\documentstyle[12pt,epsf,psfig,graphicx]{article}
\begin{document}
\title{{\bf Two stories outside Boltzmann-Gibbs statistics: Mori's }$q${\bf -phase
transitions and glassy dynamics at the onset of chaos}}
\author{A. Robledo$^{1}$\thanks{%
email: robledo@fisica.unam.mx}, F. Baldovin$^{1,2}$\thanks{%
email: baldovin@pds.infn.it} and E. Mayoral$^{1,}$\thanks{%
email: estela@eros.pquim.unam.mx} \\
%EndAName$^{1}$Instituto de F\'{i}sica,
Universidad Nacional Aut\'{o}noma de M\'{e}xico,\\
Apartado Postal 20-364, M\'{e}xico 01000 D.F., Mexico\\
$^{2}$Dipartimento de Fisica,\\
Universit\`{a} di Padova,
Via Marzolo 8, I-35131 Padova, Italy}
\date{.}
\maketitle
\begin{abstract}
First, we analyze trajectories inside the Feigenbaum attractor and obtain
the atypical weak sensitivity to initial conditions and loss of information
associated to their dynamics. We identify the Mori singularities in its
Lyapunov spectrum with the appearance of a special value for the entropic
index $q$ of the Tsallis statistics. Secondly, the dynamics of iterates at
the noise-perturbed transition to chaos is shown to exhibit the
characteristic elements of the glass transition, e.g. two-step relaxation,
aging, subdiffusion and arrest. The properties of the bifurcation gap
induced by the noise are seen to be comparable to those of a supercooled
liquid above a glass transition temperature.

\end{abstract}

\section{Introduction}

Evidence for the incidence of nonextensive dynamical properties at critical
attractors in low dimensional nonlinear maps has accumulated and advanced
over the last few years; specially with regards to the onset of chaos in
logistic maps - the Feigenbaum attractor \cite{tsallis2,robmori1},
and at the accompanying pitchfork and tangent bifurcations \cite{robledo1,baldovin3}. The more general chaotic attractors with positive Lyapunov
coefficients have full-grown phase-space ergodic and mixing properties, and
their dynamics is compatible with the Boltzmann-Gibbs (BG) statistics. As a
difference, critical attractors have vanishing Lyapunov coefficients,
exhibit memory-retentive nonmixing properties, and are therefore to be
considered outside BG statistics.

Naturally, some basic questions about the understanding of the dynamics at
critical attractors are of current interest. We mention the following: Why
do the anomalous sensitivity to initial conditions $\xi _{t}$ and its
matching Pesin identity obey the expressions suggested by the nonextensive
formalism? How does the value of the entropic index $q$ arise? Or is there a
preferred set of $q$ values? Does this index, or indexes, point to some
specific observable properties at the critical attractor?

From a broader point of view it is of interest to know if the anomalous
dynamics found for critical attractors bears some correlation with the
dynamical behavior at extremal or transitional states in systems with many
degrees of freedom. Two specific suggestions have been recently advanced, in
one case the dynamics at the onset of chaos has been demonstrated to be
closely analogous to the glassy dynamics observed in supercooled molecular
liquids \cite{robglass1}, and in the second case the dynamics at the tangent
bifurcation has been shown to be related to that at thermal critical states \cite{robcrit1}.

With regard to the above comments here we briefly recount the following
developments:

i) The finding \cite{robmori1} that the dynamics at the onset of chaos is
made up of an infinite family of Mori's $q$-phase transitions \cite{mori1,mori2}, each associated to orbits that have common starting and
finishing positions located at specific regions of the attractor. Every one
of these transitions is related to a discontinuity in the $\sigma $ function
of 'diameter ratios' \cite{schuster1}, and this in turn implies a $q$
-exponential $\xi _{t}$ and a spectrum of $q$-Lyapunov coefficients equal to
the Tsallis rate of entropy production for each set of attractor regions.
The transitions come in pairs with conjugate indexes $q$ and $Q=2-q$, as
these correspond to switching starting and finishing orbital positions. The
amplitude of the discontinuities in $\sigma $ diminishes rapidly and
consideration only of its dominant one, associated to the most crowded and
sparse regions of the attractor, provides a very reasonable description of
the dynamics, consistent with that found in earlier studies \cite{tsallis2,baldovin2}.

ii) The realization \cite{robglass1} that the dynamics at the
noise-perturbed edge of chaos in logistic maps is analogous to that observed
in supercooled liquids close to vitrification. Four major features of glassy
dynamics in structural glass formers, two-step relaxation, aging, a
relationship between relaxation time and configurational entropy, and
evolution from diffusive to subdiffusive behavior and finally arrest, are
shown to be displayed by the properties of orbits with vanishing Lyapunov
coefficient. The previously known properties in control-parameter space of
the noise-induced bifurcation gap \cite{schuster1,crutchfield1} play
a central role in determining the characteristics of dynamical relaxation at
the chaos threshold.

\section{Mori's $q$-phase transitions at onset of chaos}

The dynamics at the chaos threshold $\mu =\mu _{c}$ of the $z$-logistic map 
\begin{equation}
f_{\mu }(x)=1-\mu \left| x\right| ^{z},\;z>1,-1\leq x\leq 1,
\end{equation}
has been analyzed recently \cite{baldovin1,robmori1}. The orbit with
initial condition $x_{0}=0$ (or equivalently, $x_{0}=1$) consists of
positions ordered as intertwined power laws that asymptotically reproduce
the entire period-doubling cascade that occurs for $\mu <\mu _{c}$. This
orbit is the last of the so-called 'superstable' periodic orbits at $%
\overline{\mu }_{n}<$ $\mu _{c}$, $n=1,2,...$ \cite{schuster1}, a superstable
orbit of period $2^{\infty }$. There, the ordinary Lyapunov coefficient $%
\lambda _{1}$ vanishes and instead a spectrum of $q$-Lyapunov coefficients $%
\lambda _{q}^{(k)}$ develops. This spectrum originally studied in Refs. \cite{mori2} when $z=2$, has been shown \cite{baldovin2,robmori1} to be
associated to a sensitivity to initial conditions $\xi _{t}$ (defined as $%
\xi _{t}(x_{0})\equiv \lim_{\Delta x_{0}\to 0}(\Delta x_{t}/\Delta x_{0})$
where $\Delta x_{0}$ is the initial separation of two orbits and $\Delta
x_{t}$ that at time $t$) that obeys the $q$-exponential form 
\begin{equation}
\xi _{t}(x_{0})=\exp _{q}[\lambda _{q}(x_{0})t]\equiv [1-(q-1)\lambda
_{q}(x_{0})\ t]^{-1/q-1}
\end{equation}
suggested by the Tsallis statistics. Notably, the appearance of a specific
value for the $q$ index (and actually also that for its conjugate value $%
Q=2-q$) works out \cite{robmori1} to be due to the occurrence of Mori's '$q$%
-phase transitions' \cite{mori1} between 'local attractor structures' at $%
\mu _{c}$.

\begin{figure}[htb]
\setlength{\abovecaptionskip}{0pt} 
\centering
\includegraphics[width=9cm 
,angle=-90]{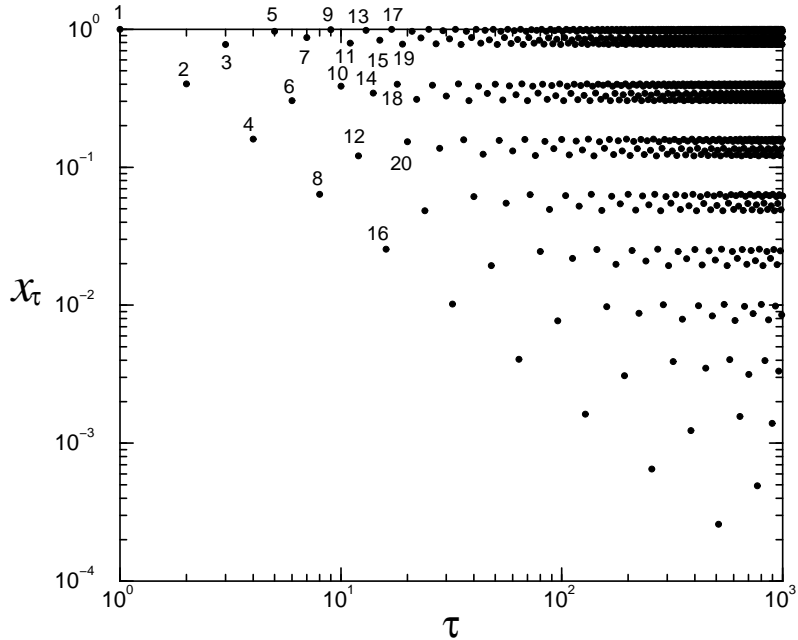}
\caption{Absolute values of positions in logarithmic scales of
iterations $\tau $ for a trajectory at }$\mu _{c}${\small \ with initial
condition $x_{0}=0$. The numbers correspond to iteration times.}
\label{fig1}
\end{figure}

As shown in Fig. 1, the absolute values for the positions $x_{\tau }$ of the
trajectory with $x_{t=0}=0$ at time-shifted $\tau =t+1$ have a structure
consisting of subsequences with a common power-law decay of the form $\tau
^{-1/1-q}$ with $q=1-\ln 2/(z-1)\ln \alpha (z)$ \cite{baldovin1}, where $%
\alpha (z)$ is the Feigenbaum universal constant that measures the
period-doubling amplification of iterate positions. That is, the attractor
can be decomposed into position subsequences generated by the time
subsequences $\tau =(2k+1)2^{n}$, each obtained by proceeding through $%
n=0,1,2,...$ for a fixed value of $k=0,1,2,...$. See Fig. 1. The $k=0$
subsequence can be written as $x_{t}=\exp _{2-q}(-\lambda _{q}^{(0)}t)$ with 
$\lambda _{q}^{(0)}=(z-1)\ln \alpha (z)/\ln 2$.

$q${\it -Lyapunov coefficients}. The sensitivity $\xi _{t}(x_{0})$ can be
obtained \cite{robmori1} from $\xi _{t}(m)\simeq \left| \sigma
_{n}(m-1)/\sigma _{n}(m)\right| ^{n}$, $t=2^{n}-1$, $n\ $large, where $%
\sigma _{n}(m)=d_{n+1,m}/d_{n,m}$ and where $d_{n,m}$ are the diameters that
measure adjacent position distances that form the period-doubling cascade
sequence \cite{schuster1}. Above, the choices $\Delta x_{0}=d_{n,m}$ and $%
\Delta x_{t}=d_{n,m+t}$, $t=2^{n}-1$, have been made for the initial and the
final separation of the trajectories, respectively. In the large $n$ limit $%
\sigma _{n}(m)$ develops discontinuities at each rational $m/2^{n+1}$ \cite
{schuster1}, and according to our expression for $\xi _{t}(m)$ the
sensitivity is determined by these discontinuities. For each discontinuity
of $\sigma _{n}(m)$ the sensitivity can be written in the forms $\xi
_{t}=\exp _{q}[\lambda _{q}t]$ and $\xi _{t}=\exp _{2-q}[\lambda _{2-q}t]$, $%
\lambda _{q}>0$ and $\lambda _{2-q}<0$ \cite{robmori1}. This result reflects
the multi-region nature of the multifractal attractor and the memory
retention of these regions in the dynamics. The pair of $q$-exponentials
correspond to a departing position in one region and arrival at a different
region and vice versa, the trajectories expand in one sense and contract in
the other. The largest discontinuity of $\sigma _{n}(m)$ at $m=0$ is
associated to trajectories that start and finish at the most crowded ($%
x\simeq 1$) and the most sparse ($x\simeq 0$) regions of the attractor. In
this case one obtains 
\begin{equation}
\lambda _{q}^{(k)}=\frac{(z-1)\ln \alpha (z)}{(2k+1)\ln 2}>0,\ k=0,1,2,...,
\end{equation}
the positive branch of the Lyapunov spectrum, when the trajectories start at 
$x\simeq 1$ and finish at $x\simeq 0$. By inverting the situation one
obtains 
\begin{equation}
\lambda _{Q}^{(k)}=-\frac{2(z-1)\ln \alpha (z)}{(2k+1)\ln 2}<0,\ k=0,1,2,...,
\end{equation}
the negative branch of the Lyapunov spectrum. Notice that $\exp
_{2-q}(y)=1/\exp _{q}(-y)$. So, when considering these two dominant families
of orbits all the $q$-Lyapunov coefficients appear associated to only two
specific values of the Tsallis index, $q$ and $Q=2-q$.

{\it Mori's }$q${\it -phase transitions}. As a function of the running
variable $-\infty <{\sf q}<\infty $ the $q$-Lyapunov coefficients become a
function $\lambda ({\sf q})$ with two steps located at ${\sf q}=q=1-\ln
2/(z-1)\ln \alpha (z)$ and ${\sf q}=Q=2-q$. In this manner contact can be
established with the formalism developed by Mori and coworkers \cite{mori1}
and the $q$-phase transition obtained in Refs. \cite{mori2}. The step
function for $\lambda ({\sf q})$ can be integrated to obtain the spectrum $%
\phi ({\sf q})$ ($\lambda ({\sf q})\equiv d\phi /d\lambda ({\sf q})$) and
its Legendre transform $\psi (\lambda )$ ($\equiv \phi -(1-{\sf q})\lambda $%
), the dynamic counterparts of the Renyi dimensions $D({\sf q})$ and the
spectrum $f(\widetilde{\alpha })$ that characterize the geometry of the
attractor. The result for $\psi (\lambda )$ is
\begin{equation}
\psi (\lambda )=\left\{ 
\begin{array}{l}
(1-Q)\lambda ,\ \lambda _{Q}^{(0)}<\lambda <0, \\ 
(1-q)\lambda ,\ 0<\lambda <\lambda _{q}^{(0)}.
\end{array}
\right. 
\end{equation}
As with ordinary thermal 1st order phase transitions, a ''$q$-phase''
transition is indicated by a section of linear slope $m=1-q$ in the spectrum
(free energy) $\psi (\lambda )$, a discontinuity at $q$ in the Lyapunov
function (order parameter) $\lambda ({\sf q})$, and a divergence at $q$ in
the variance (susceptibility) $v({\sf q})$. For the onset of chaos at $\mu
_{c}(z=2)$ a $q$-phase transition was numerically determined \cite{mori1, mori2}. According to $\psi (\lambda )$ above we obtain a conjugate pair
of $q$-phase transitions that correspond to trajectories linking two regions
of the attractor, the most crowded and most sparse. See Fig. 2. Details
appear in Ref. \cite{robmori1}.

\begin{figure}[htb]
\setlength{\abovecaptionskip}{0pt} 
\centering
\includegraphics[width=9cm 
,angle=0]{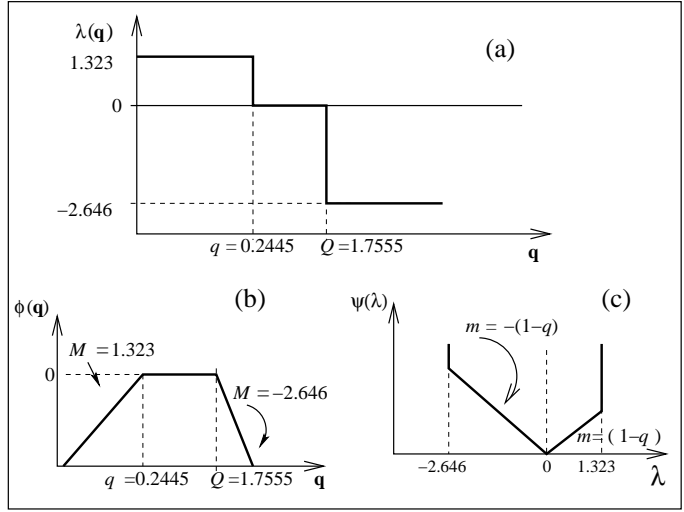}
\caption{$q$-phase transitions with index values $q=0.2445$ and $Q=2-q=1.7555$ obtained for $z=2$ from the main
discontinuity in $\sigma _{n}(m)$. See text for details.}
\label{fig2}
\end{figure}

{\it Generalized Pesin identity}. Ensembles of trajectories with starting
points close to the attractor point $x_{0}$ expand in such a way that a
uniform distribution of initial conditions remains uniform for all later
times $t$. As a consequence of this we established \cite{baldovin2,robmori1} the identity of the rate of entropy production $K_{q}^{(k)}$ with 
$\lambda _{q}^{(k)}$. The $q$-generalized rate of entropy production $K_{q}$
is defined via $K_{q}t=S_{q}(t)-S_{q}(0)$, $t$ large, where 
\begin{equation}
S_{q}\equiv \sum_{i}p_{i}\ln _{q}\left( \frac{1}{p_{i}}\right) =\frac{%
1-\sum_{i}^{W}p_{i}^{q}}{q-1}
\end{equation}
is the Tsallis entropy, $p_{i}$ is the trajectories' distribution, and where 
$\ln _{q}y\equiv (y^{1-q}-1)/(1-q)$ is the inverse of $\exp _{q}(y)$. See
Figs. 2 and 3 in Ref. \cite{baldovin2}.

\section{Glassy dynamics at noise-perturbed onset of chaos}

We describe now the effect of additive noise in the dynamics at the onset of
chaos. The logistic map $z$ $=2$ reads now 
\begin{equation}
x_{t+1}=f_{\mu }(x_{t})=1-\mu x_{t}^{2}+\chi _{t}\sigma ,\ -1\leq x_{t}\leq
1,\ 0\leq \mu \leq 2,
\end{equation}

where $\chi _{t}$ is Gaussian-distributed with average $\left\langle \chi
_{t}\chi _{t^{\prime }}\right\rangle =\delta _{t.t^{\prime }}$, and $\sigma $
is the noise intensity. For $\sigma >0$ the noise fluctuations wipe the fine
features of the periodic attractors as these widen into bands similar to
those in the chaotic attractors, nevertheless there remains a well-defined
transition to chaos at $\mu _{c}(\sigma )$ where the Lyapunov exponent $%
\lambda _{1}$ changes sign. The period doubling of bands ends at a finite
maximum period $2^{N(\sigma )}$ as $\mu \rightarrow \mu _{c}(\sigma )$ and
then decreases at the other side of the transition. This effect displays
scaling features and is referred to as the bifurcation gap \cite{schuster1,crutchfield1}. When $\sigma >0$ the trajectories visit sequentially a
set of $2^{n}$ disjoint bands or segments leading to a cycle, but the
behavior inside each band is fully chaotic. These trajectories represent
ergodic states as the accessible positions have a fractal dimension equal to
the dimension of phase space. When $\sigma =0$ the trajectories correspond
to a nonergodic state, since as $t\rightarrow \infty $ the positions form
only a Cantor set of fractal dimension $d_{f}=0.5338...$. Thus the removal
of the noise $\sigma \rightarrow 0$ leads to an ergodic to nonergodic
transition in the map.

As shown in Ref. \cite{robglass1} when $\mu _{c}(\sigma >0)$ there is a
'crossover' or 'relaxation' time $\tau _{x}=\sigma ^{r-1}$, $r\simeq 0.6332$%
, between two different time evolution regimes. This crossover occurs when
the noise fluctuations begin suppressing the fine structure of the attractor
as displayed by the superstable orbit with $x_{0}=0$ described previously.
For $\tau <\tau _{x}$ the fluctuations are smaller than the distances
between the neighboring subsequence positions of the $x_{0}=0$ orbit at $\mu
_{c}(0)$, and the iterate position with $\sigma >0$ falls within a small
band around the $\sigma =0$ position for that $\tau $. The bands for
successive times do not overlap. Time evolution follows a subsequence
pattern close to that in the noiseless case. When $\tau \sim \tau _{x}$ the
width of the noise-generated band reached at time $\tau _{x}=2^{N(\sigma )}$
matches the distance between adjacent positions, and this implies a cutoff
in the progress along the position subsequences. At longer times $\tau >\tau
_{x}$ the orbits no longer trace the precise period-doubling structure of
the attractor. The iterates now follow increasingly chaotic trajectories as
bands merge with time. This is the dynamical image - observed along the time
evolution for the orbits of a single state $\mu _{c}(\sigma )$ - of the
static bifurcation gap initially described in terms of the variation of the
control parameter $\mu $ \cite{crutchfield1}.

{\it Two-step relaxation}. Amongst the main dynamical properties displayed
by supercooled liquids on approach to glass formation is the growth of a
plateau, and for that reason a two-step process of relaxation, in the time
evolution of two-time correlations \cite{debenedetti1}. \begin{figure}[btb]
\setlength{\abovecaptionskip}{0pt} 
\centering
\includegraphics[width=9cm 
,angle=0]{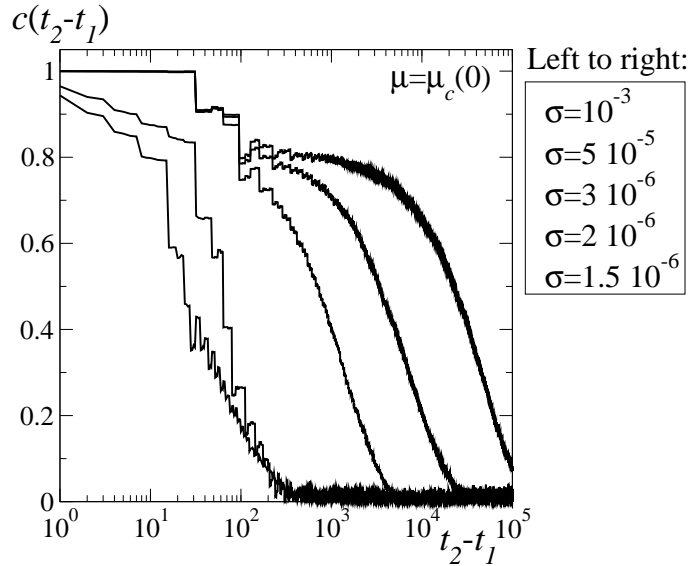}
\caption{Two-time correlation function $c(t_{2}-t_{1})$ for
an ensemble of trajectories with $x_{0}=0$ for different values of
noise amplitude $\sigma $. See text for details.}
\label{fig3}
\end{figure}
This consists of a primary power-law decay in time difference $\Delta t$ (so-called $\beta $
relaxation) that leads into the plateau, the duration $t_{x}=\tau _{x}-1$ of
which diverges also as a power law of the difference $T-T_{g}$ as the
temperature $T$ decreases to a glass temperature $T_{g}$. After $t_{x}$
there is a secondary power law decay (so-called $\alpha $ relaxation) away
from the plateau \cite{debenedetti1}. In Fig. 3 we show \cite{baldovin4} the behavior of the correlation function 
\begin{equation}
c(t_{2}-t_{1})=\frac{\left\langle x_{t_{2}}x_{t_{1}}\right\rangle
-\left\langle x_{t_{2}}\right\rangle \left\langle x_{t_{1}}\right\rangle }{%
\chi _{t_{1}}\chi _{t_{2}}},
\end{equation}
for different values of noise amplitude. Above, $\left\langle
...\right\rangle $ represents an average over an ensemble of trajectories
starting at $x_{0}=0$ and $\chi _{t_{i}}=\sqrt{\left\langle
x_{t_{i}}^{2}\right\rangle -\left\langle x_{t_{i}}\right\rangle ^{2}}$. The
development of the two power-law relaxation regimes and their intermediate
plateau can be clearly appreciated. See Ref. \cite{robglass1} for the
interpretation of the map analogs of the $\alpha $ and $\beta $ relaxation
processes.

{\it Aging scaling.} A second important (nonequilibrium) dynamical property
of glasses is the loss of time translation invariance observed for $T\leq T_{g}$, a characteristic known as aging. The drop time of
relaxation functions and correlations display a scaling dependence on the
ratio $t/t_{w}$ where $t_{w}$ is a waiting time. In Fig. 4a we show \cite{baldovin4} the correlation function 
\begin{figure}[htb]
\setlength{\abovecaptionskip}{0pt} 
\centering
\includegraphics[width=6cm 
,angle=0]{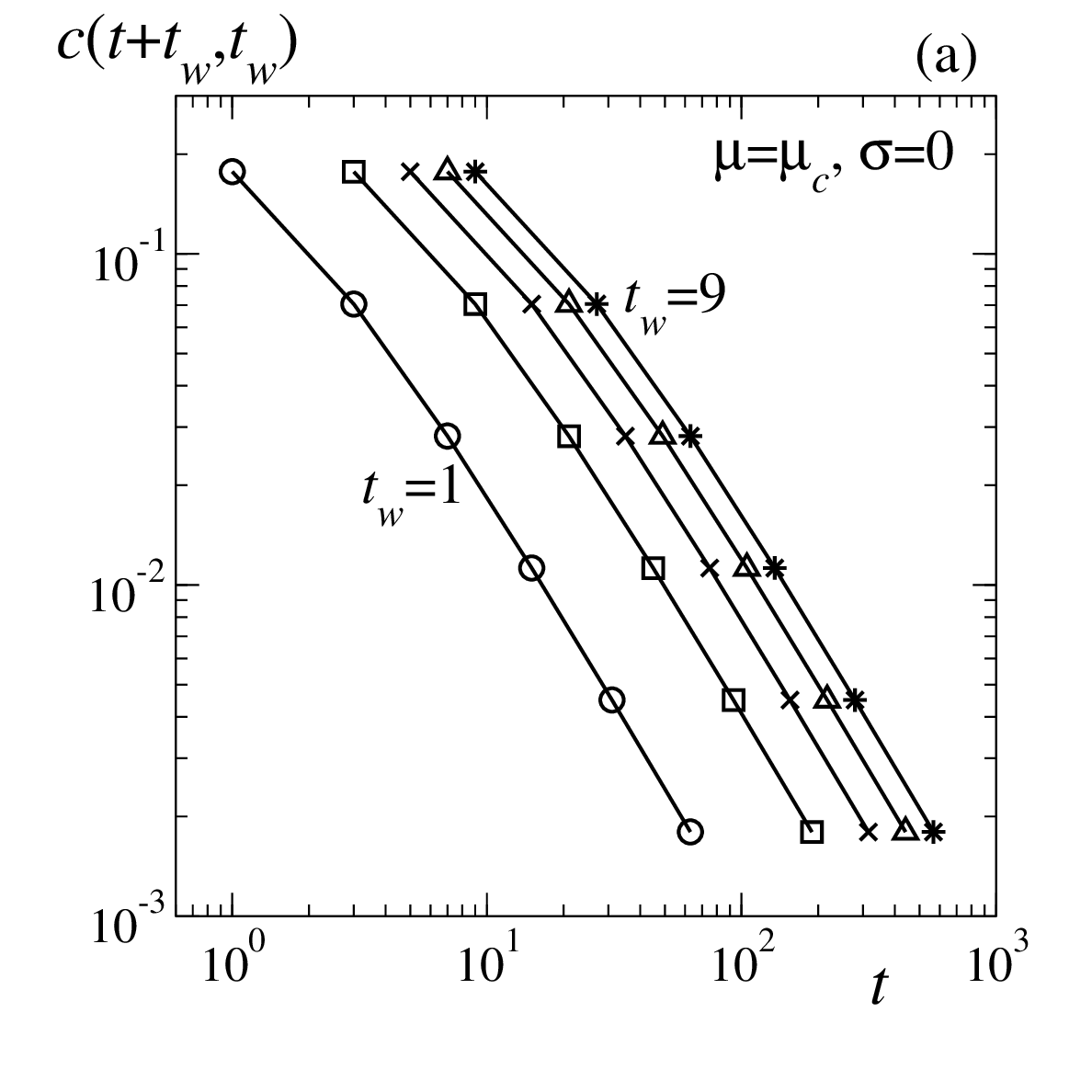}
\includegraphics[width=6cm 
,angle=0]{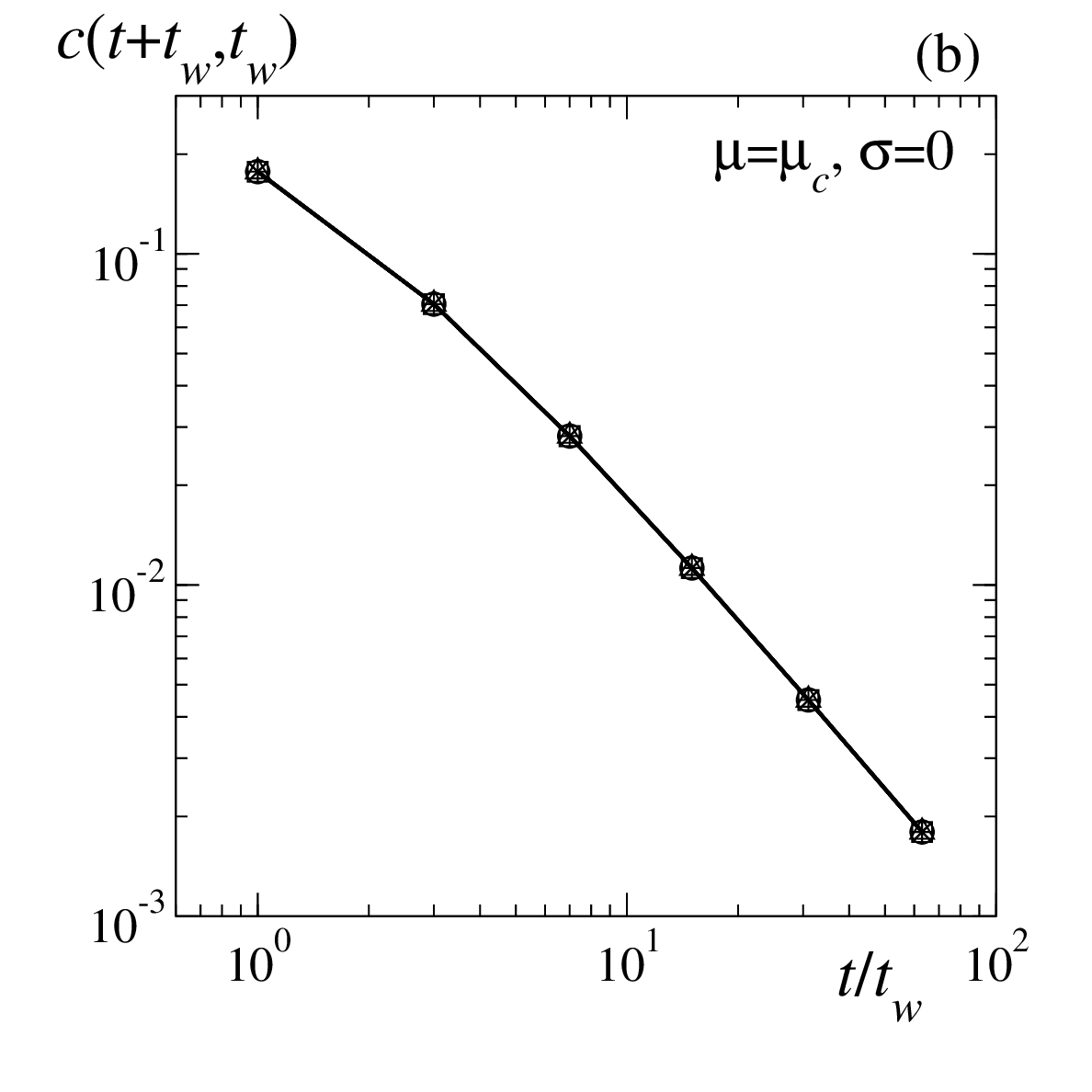}
\caption{a) Two-time correlation function }$c(t+t_{w},t_{w})$
for different values of $\sigma $. b) The same data in terms of the
rescaled variable $t/t_{w}.$ See text for details.]
\label{fig4}
\end{figure}
\begin{equation}
c(t+t_{w},t_{w})=(1/N)\sum_{j=1}^{N}x_{(t+t_{w})j}x_{tj}
\end{equation}
for different values of $\sigma $, and in Fig. 4b the same data where the
rescaled variable $t/t_{w}=2^{n}-1$, $t_{w}=2k+1$, $k=0,1,...$, has been
used. The characteristic aging scaling behavior is patent. See Ref. \cite{robglass1} for an analytical description of the built-in aging properties
of the trajectories at $\mu _{c}(\sigma )$.

{\it Adam-Gibbs relation}. A third notable property is that the
experimentally observed relaxation behavior of supercooled liquids is well
described, via standard heat capacity assumptions \cite{debenedetti1}, by the
so-called Adam-Gibbs equation, $t_{x}=A\exp (B/TS_{c})$, where $t_{x}$ is
the relaxation time at $T$, and the configurational entropy $S_{c}$ is
related to the number of minima of the fluid's potential energy surface \cite
{debenedetti1}. See Ref. \cite{robglass1} for the derivation of the analog
expression for the nonlinear map. Instead of the exponential Adam-Gibbs
equation, this expression turned out to have the power law form 
\begin{equation}
t_{x}=(s/S_{c})^{(1-r)/r}.
\end{equation}
Since $(1-r)/r\simeq 0.5792$ then $t_{x}\rightarrow \infty $ and $%
S_{c}\rightarrow 0$ as $\sigma \rightarrow 0$.

{\it Subdiffusion and arrest}. A fourth distinctive property of supercooled
liquids on approach to vitrification is the progression from normal
diffusivity to subdiffusive behavior and finally to a halt in the growth of
the molecular mean square displacement.
\begin{figure}[htb]
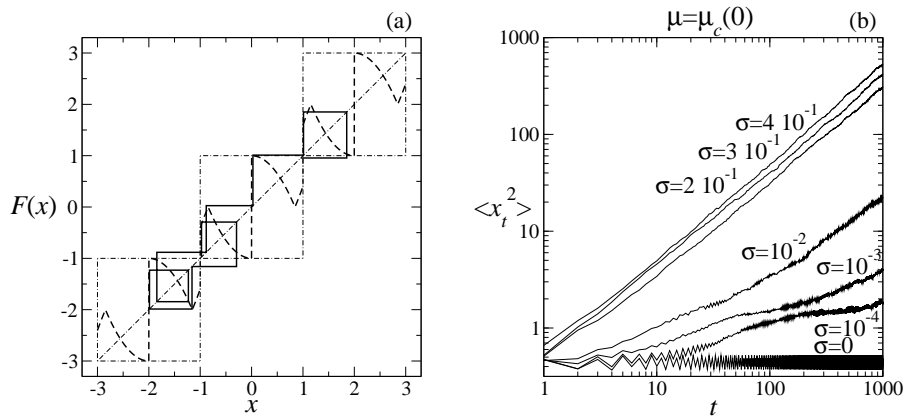

\setlength{\abovecaptionskip}{0pt} 
\centering
\includegraphics[width=6cm 
,angle=0]{diff_cell.eps}
\includegraphics[width=6cm 
,angle=0]{diff_ensemble.eps}
\caption{a) Repeated-cell map and trajectory. b) Mean square
displacement $\left\langle x_{t}^{2}\right\rangle $ for
trajectories with $x_{0}=0$ for several values of noise amplitude
$\sigma $. See text for details.}
\label{fig5}
\end{figure}
 To investigate this aspect of
vitrification in the map at $\mu _{c}(\sigma )$, we constructed \cite{baldovin4} a periodic map with repeated cells of the form $x_{t+1}=F(x_{t})$%
, $F(l+x)=l+F(x)$, $l=...-1,0,1,...$, $F(-x)=F(x)$, where

\begin{equation}
F(x)=\left\{ 
\begin{array}{c}
-\left| 1-\mu _{c}x^{2}\right| +\chi \sigma ,\;-1\leq x<0, \\ 
\left| 1-\mu _{c}x^{2}\right| +\chi \sigma ,\;0\leq x<1.
\end{array}
\right. 
\end{equation}

Fig. 5a shows this map together with a portion of one of its trajectories,
while Fig. 5b shows the mean square displacement $\left\langle
x_{t}^{2}\right\rangle $ as obtained from an ensemble of trajectories with $%
x_{0}=0$ for several values of noise amplitude. The progression from normal
diffusion to subdiffusion and to final arrest can be plainly observed as $%
\sigma \rightarrow 0$ \cite{baldovin4}.

\section{Summary}

We reviewed recent understanding on the dynamics at the onset of chaos in
the logistic map. We exhibited links between previous developments, such as
Feigenbaum's $\sigma $ function, Mori's $q$-phase transitions and the
noise-induced bifurcation gap, with more recent advances, such as $q$%
-exponential sensitivity to initial conditions \cite{baldovin1,mayoral1}, $q$-generalized Pesin identity \cite{baldovin2,robmori1} and dynamics of glass formation \cite{robglass1}.

An important finding is that the dynamics is constituted by an infinite
family of Mori's $q$-phase transitions, each associated to orbits that have
common starting and finishing positions located at specific regions of the
attractor. Thus, the special values for the Tsallis entropic index $q$ in $%
\xi _{t}$ and $S_{q}$ are equal to the special values of the variable $q$ at
which the $q$-phase transitions take place.

As described, the dynamics of noise-perturbed logistic maps at the chaos
threshold presents the characteristic features of glassy dynamics observed
in supercooled liquids. The limit of vanishing noise amplitude $\sigma
\rightarrow 0$ (the counterpart of the limit $T-T_{g}\rightarrow 0$ in the
supercooled liquid) leads to loss of ergodicity. This nonergodic state with $%
\lambda _{1}=0$ corresponds to the limiting state, $\sigma \rightarrow 0$, $%
t_{x}\rightarrow \infty $, of a family of small $\sigma $ states with glassy
properties, which are expressed for $t<t_{x}$ via the $q$-exponentials of
the Tsallis formalism.

Acknowledgments. FB warmly acknowledges hospitality at UNAM where part of
this work has been done. Work partially supported by DGAPA-UNAM and CONACyT
(Mexican Agencies).

\end{document}